\documentclass[11pt,a4paper,twoside, groupcitations]{article}
\usepackage[T1]{fontenc}
\usepackage[ansinew]{inputenc}
\usepackage[english]{babel}
\usepackage{amsfonts}
\usepackage{amsmath}
\usepackage{bm}
\usepackage{array}
\usepackage{amsthm}
\usepackage{amssymb}
\usepackage{graphicx}
\usepackage{subfigure}
\usepackage{braket}
\usepackage{eucal}
\usepackage{verbatim}
\usepackage[table]{xcolor}
\usepackage{caption}
\usepackage{cite}
\usepackage{textcomp}
\usepackage{multicol}
\usepackage{tikz}
\usetikzlibrary{positioning,arrows}
\usetikzlibrary{decorations.pathmorphing}
\usetikzlibrary{decorations.markings}
\usetikzlibrary{calc,decorations.markings}
\usetikzlibrary{arrows,shapes}
\usetikzlibrary{matrix,arrows}
\usepackage{pgfplots}
\usepackage{xparse}
\definecolor{jade}{HTML}{00A86B}
\newcommand{\be}{\begin{eqnarray}}
\newcommand{\ee}{\end{eqnarray}}
\renewcommand{\d}{\mbox{${\rm d}$}} %d differenziale non corsivo in math mode

\newcommand{\gn}{G_{\rm N}}
\newcommand{\rh}{r_{\rm H}}

\def\beq{\begin{equation}}
\def\eeq{\end{equation}}
\def\@fmsl@sh#1#2#3{\m@th\ooalign{$\hfil#1\mkern#2/\hfil$\crcr$#1#3$}}
 \def\eq#1\en{\begin{equation}#1\end{equation}}
\def\s[#1,#2]{[#1\stackrel{\star}{,}#2]}
\def\sx[#1,#2]{[#1\stackrel{\star_{x}}{,}#2]}

\def\beq{\begin{equation}}
\def\eeq{\end{equation}}

\title{\bf Dynamical stability of bootstrapped Newtonian stars}
\author{Octavian Micu\thanks{E-mail: octavian.micu@spacescience.ro}
\\
\\
{\em Institute of Space Science - Subsidiary of INFLPR}
\\
{\em P.O. Box MG-23, RO-077125 Bucharest-Magurele, Romania}
\\
\\
}
\begin{document}
\maketitle
\begin{abstract}
We investigate the dynamical stability of bootstrapped Newtonian stars following homologous adiabatic perturbations, focusing on objects of low or intermediate compactness. 
The results show that for stars with homogeneous densities these perturbations induce some oscillatory behaviour regardless of their compactness, density and adiabatic index, which makes them dynamically stable. 
In the case or polytropes with density profiles approximated by Gaussian distributions, both stable and unstable behaviours are possible. It was also shown that in the limit in which the profile of the Gaussian density distribution flattens out, the parameter space for which the perturbations result in an oscillatory behaviour increases, which is in agreement with the case of stars with homogeneous densities. 
 
\end{abstract}  
%
%
%\pacs{}
%
%
%%%%%%%%%%%%%%%%%%%%%%%%%%%%%%%%%%%%%%%%%%%%%%%%%%%%%%%%%%%%%%%%
%%%
%%%                     INTRODUCTION
%%%
%%%%%%%%%%%%%%%%%%%%%%%%%%%%%%%%%%%%%%%%%%%%%%%%%%%%%%%%%%%%%%%%
%
\newpage
\section{Introduction}
\label{sec:intro}
\setcounter{equation}{0}
Bootstrapped Newtonian stars were discussed so far in a series of papers~\cite{Casadio:2018qeh, Casadio:2019cux, Casadio:2020kbc, Casadio:2020ueb, Casadio:2019pli, Casadio:2021gdf, Casadio:2022pme,Casadio:2022gbv}, and in this work we propose to build upon those results and look into the dynamical stability of these objects.
Bootstrapped Newtonian gravity, which can be viewed as a bottom-up extension of Newtonian gravity that includes non-linear interaction terms that have functional forms similar to those appearing at leading orders in the perturbative weak-field expansion of general relativity, is an attempt to investigate static and spherically symmetric compact sources in an intermediary framework between the two aforementioned theories in which all the additional terms are treated equivalently. The contribution of each additional term can still be tuned from the (dimensionless) coupling constants. 
One of the main achievements of this model is that it led to some interior solutions for extremely dense objects which evade the singularities that appear in the general relativistic treatment of objects collapsed behind trapping surfaces.
According to general relativity, the cores of these extremely dense objects are supposed to ultimately collapse down into a vanishing volume where the density becomes infinite. Such point-like sources are classically unacceptable, and one hopes that some quantum effects resolve this issue in the strong fields regime. Bootstrapped Newtonian sources were built on the premise that infinities such as those predicted by general relativity do not occur in nature. As it will be explained later, or can be better understood from the papers cited earlier in this paragraph, when searching for solutions for the gravitational potential, one of the constraints imposed is for the potential to be regular in the centre of the object and what is found is that such solutions are feasible. 

There are two mass parameters in the model: one is an ADM-like mass~\cite{adm} and it characterises the outer gravitational potential as it would be measured by a distant observer, while the other is the proper mass calculated by taking the volume integral of the density of the star. 
In general these two masses are different, $M\neq M_0$, regardless of the values of the other (free) parameters of the model. This was discussed in detail in Refs.~\cite{Casadio:2019pli, Casadio:2022pme} and it is a fundamental difference to Newtonian gravity stemming from the non-linear nature of the theory. 

One of the most important features of bootstrapped Newtonian stars is that no Buchdahl limit~\cite{Buchdahl:1959zz} exists for these objects. As a consequence, the pressure within these stars can be sufficient to keep their cores in equilibrium for arbitrarily large compactness values. 

After developing the model to this extent, the natural further development is to find whether bootstrapped Newtonian stars are stable to various types of perturbations. In this work we investigate their dynamical stability under the action of (radial) homologous adiabatic perturbations. In the next section we very briefly introduce the model. This is followed in Section~\ref{sec:dynamical} by a general (textbook) discussion of the dynamical stability of stars due to such adiabatic perturbations, specifically applied to bootstrapped Newtonian stars. After the general description, in Sections~\ref{sec:hom} and \ref{sec:polytropes} we apply this recipe in detail to stars with homogeneous densities, respectively to polytropic stars whose density profiles are fairly well approximated by Gaussian distributions. The last section is where we draw the conclusions. 

\section{Bootstrapped Newtonian gravity}
\setcounter{equation}{0}
\label{bsgrav}
Ever since its inception, the fundamentals of the bootstrapped Newtonian gravity model have been detailed repeatedly in several of the articles in which it was further developed. For detailed derivations the readers are directed to Refs.~\cite{Casadio:2018qeh, Casadio:2019cux, Casadio:2020kbc, Casadio:2020ueb, Casadio:2019pli, Casadio:2022pme}. 
As introduced in ~\cite{Casadio:2018qeh}, and working in units such that $c=1$, the bootstrapped Newtonian potential $V=V(r)$ for (static) spherically symmetric objects can be obtained starting from the Lagrangian 
\be 
L[V]
&\!\!=\!\!&
L_{\rm N}[V]
-4\,\pi
\int_0^\infty
r^2\,\d r
\left[
q_V\,\mathcal{J}_V\,V
+
q_p\,\mathcal{J}_p\,V
+
q_\rho\, \mathcal{J}_\rho \left(\rho+q_p\,\mathcal{J}_p\right)
\right]
\nonumber
\\
&\!\!=\!\!&
-4\,\pi
\int_0^\infty
r^2\,\d r
\left[
\frac{\left(V'\right)^2}{8\,\pi\,\gn}
\left(1-4\,q_V\, V\right)
+\left(\rho+\,q_p\,p\right)
V
\left(1-2\,q_\rho\, V\right)
\right]
\ ,
\label{LagrV}
\ee
where $f'\equiv\d f/\d r$.
In order to evidentiate the different contributions, in the above equation we separated the Lagrangian for the Newtonian potential
\be
L_{\rm N}[V]
=
-4\,\pi
\int_0^\infty
r^2 \,\d r
\left[
\frac{\left(V'\right)^2}{8\,\pi\,\gn}
+\rho\,V
\right]
\ ,
\label{LagrNewt}
\ee
which leads to the Poisson equation
\be
r^{-2}\left(r^2\,V'\right)'
\equiv
\triangle V
=
4\,\pi\,\gn\,\rho .
\label{EOMn}
\ee
Of course, in the absence of the additional terms in the Lagrangian, $V=V_{\rm N}$ stands for the Newtonian potential due to the matter density $\rho=\rho(r)$.

Each of the extra terms from Eq.~\eqref{LagrV} was discussed in more detail in the references cited in the beginning of the section. 
The first term in the integral represents the gravitational self-coupling sourced by the gravitational energy $U_{\rm N}$ per unit volume
\be
\mathcal{J}_V
\simeq
\frac{\d U_{\rm N}}{\d \mathcal{V}} 
=
-\frac{\left[ V'(r) \right]^2}{2\,\pi\,\gn}
\ .
\label{JV}
\ee
In order to be able to study all different regimes of the theory, each term couples to the potential via some coupling constant whose value varies between zero and one. For this term, the coupling is represented by $q_V$.

Before introducing the second additional term, it is worth mentioning that the bootstrapped Newtonian model was developed to be applicable in all compactness regimes, from low to high compactness values. The compactness is defined in terms of the ADM-like mass $M$, which represents the mass that an observer would calculate for the star when studying orbits around it~\cite{adm,Casadio:2021gdf,DAddio:2021xsu}, and the star radius $R$ as follows
\be
X
\equiv
\frac{\gn\, M}{R} \ .
\label{defX}
\ee

In the high compactness regime, meaning for $X\gtrsim 1$, the static pressure $p=p(r)$ becomes very large~\cite{Casadio:2018qeh}, which prompted the addition of a potential energy term $U_p$ such that
\be
\mathcal{J}_p
\simeq
-\frac{\d U_p}{\d \mathcal{V}} 
=
p
\ ,
\label{JP}
\ee
and this term couples to the potential via the positive constant $q_p$. This second term just adds to the density $\rho$, and its effect is only to shift $\rho \to \rho+q_p\,p$, where we can make use of the constant $q_p$ to implement the non-relativistic limit as $q_p\to 0$. This shift of the density to include the pressure should remind the readers of the definition of the Tolman mass~\cite{tolman}.

The total Lagrangian in Eq.~\eqref{LagrV} is finally completed by an additional higher-order term which couples with the matter source
\be
\mathcal{J}_\rho=-2\,V^2
\ .
\ee
\par
Starting from Eq.~\eqref{LagrV} it is a simple exercise to obtain the Euler-Lagrange equation for the potential
\be
\triangle V
=
4\,\pi\,\gn\left(\rho+q_p\,p\right)
\frac{1-4\,q_\rho\,V}{1-4\,q_V\,V}
+
\frac{2\,q_V\left(V'\right)^2}
{1-4\,q_V\,V}
\ .
\label{EOMV}
\ee

The (dimensionless) couplings $q_V$, $q_p$ and $q_\rho$ allow one to see the effects of each additional term and their different values can be related to specific theories of the gravity-matter interaction.
In the limit $q_V=q_p=q_\rho\to 0$ we can easily see that the Newtonian limit is obtained, since the only remaining term in~Eq.~\eqref{LagrV} is the Lagrangian for the Newtonian potential.

In addition to the Euler-Lagrange equation, the system is constrained by the conservation equation which determines the pressure
\be
p'
=
-\left(\rho+q_p\,p\right) V'
\ . 
\label{eqP}
\ee 
This form of the conservation equation includes a correction to the Newtonian formula which is necessary in order to include the (non-negligible) contribution of the pressure to the energy density. Another interpretation for this is as an approximation of the general relativistic Tolman-Oppenheimer-Volkoff equation.

After this very brief introduction to the model in which all couplings were specified, for the remainder of the paper we will assume all their values to be $q_V=q_p=q_\rho=1$. This is the regime in which all terms contribute the most, given that all couplings are set to their maximum value. 
In a future work it will be interesting to investigate different regimes, in which the contributions of the additional terms of the Lagrangian from Eq.~\eqref{LagrV} are tuned using the coupling constants, as it was done in~\cite{Casadio:2019pli}. However, such an exhaustive analysis is beyond the scope of this paper.
This simplification reduces Eq.~\eqref{EOMV} to
\be
\triangle V
=
4\,\pi\,\gn\left(\rho+p\right)
+
\frac{2\,\left(V'\right)^2}
{1-4\,\,V}
\ .
\label{EOMVq1}
\ee

\section{Dynamical stability of stars}
\label{sec:dynamical}
\setcounter{equation}{0}

Stars are maintained in hydrostatic equilibrium by the balance between the pressure which counteracts the inwards oriented gravitational force.
This means that a thin spherical shell of thickness $\d r$ from inside the star is subjected to two forces. First there's the gravitational force (per unit area) which, for a spherically symmetric object, is given by 
\be
F_{g}= -\left(\rho+p\right)\, V' \, \d r \ , 
\ee
where, additionally to the purely Newtonian case, the pressure is also considered to be a source for gravitation, as it was detailed in the previous section. Note that we used $q_p=1$.
The pressure force per unit area that acts on this shell in the limit in which the thickness of the shell $\d r \rightarrow 0$ is 
\be
F_p&=&p(r)-p\,(r\,+\,\d r) \nonumber
\\
&=& - p' \, \d r \ .
\ee
With these, we can write Newton's second law of motion for the shell of mass per unit area $\rho\, \d r$
\be
(\rho\, \d r)\, \ddot{r} = -\left[\left(\rho\,+p\right)\, V'+ p' \right] \d r \ ,
\ee
or 
\be
\ddot{r} = -\frac{\rho\,+p}{\rho}\, V'- \frac{1}{\rho}\, p' \ ,
\label{acc}
\ee
where $\dot{f} \equiv \d f/\d t$. 
This equation governs the motion of the shell in response to departures from equilibrium. The star is in hydrostatic equilibrium when the acceleration from the left hand side of Eq.~\eqref{acc} is null, case in which the remaining terms can be written as
\be
p'= - \left(\rho\,+p\right) V' \ ,
\ee
which, as expected, is the same as Eq.~\eqref{eqP} with $q_p=1$. Moreover, in the limit $p\rightarrow 0$, the Newtonian case is recovered. To be even more precise, the Newtonian limit is also obtained directly from~\eqref{eqP} by setting $q_p=0$.

\subsection{Adiabatic perturbations}

An important step in further developing the model is understanding how dynamical instabilities, which are the result of departures from hydrostatic equilibrium, affect bootstrapped Newtonian stars. 
The investigation is performed under the assumption that the stars behave adiabatically (in this simple model we neglect any heat exchange), which means that the pressure and density are related by the equation for adiabatic gasses 
\be
p=p_0\,\left(\frac{\rho}{\rho_0}\right)^{\gamma} \ ,
\label{adiabatic}
\ee
where $\gamma$ represents the adiabatic index, respectively $p_0$ and $\rho_0$ are the unperturbed pressure and density. In general both $p_0$ and $\rho_0$ depend (implicitly) on the radius and for any shell of radius $0\leq r_0\leq R$ one can write the hydrostatic equilibrium equation
\be
\left(1+ \frac{p_0}{\rho_0}\right)\, V'(r_0) + \frac{1}{\rho_0}\, p_0' = 0 \ .
\label{acczero}
\ee

We are interested in global instabilities such as those induced by homologous perturbations, which are perturbations that expand or contract the star in a uniform manner, in the sense that if the star is divided in thin spherical concentrical shells of thickness $\d r$, all these shells expand or shrink by the same (small) amount $\delta r$. 
For instance, the shell of matter located initially at radius $r_0$ will shift to radius 
$r_0\,(1+\delta r/r_0)$, where $\delta r/r_0\ll 1$. A positive sign for $\delta r$ means that the star is expanding, while for a negative sign of the same quantity the star is contracting. 

During this process, the proper mass $\d m_0 $ of the shell is assumed to remain constant. This means that the perturbation affects the density of the star which changes from $\rho_0$ to $\rho_0\,(1+\delta \rho/\rho_0)$, where the perturbation of the density is also very small $\delta \rho/\rho_0\ll 1$. Since the density of the star is inversly proportional to the volume, one can easily show that at the lowest order the change in density and the change in radius are related by 
\be
\frac{\delta \rho}{\rho_0} 
=
-3\,\frac{\delta r}{r_0} \ . 
\label{delrho} 
\ee 

The pressure within the star is also affected by the perturbation. Both the perturbed pressure and density and their corresponding unperturbed values must satisfy the same adiabatic equation
\be
p_0\left(1+\frac{\delta p}{p_0} \right)=\frac{p_0}{\rho_0^\gamma} \left[ \rho_0\left(1+\frac{\delta \rho}{\rho_0} \right)\right]^{\gamma} \ ,
\ee
from where, after Taylor expanding the right hand side and keeping the lowest order term in $\delta \rho/\rho_0$, we find
\be
\frac{\delta p}{p_0} 
=
\gamma \,\frac{\delta \rho}{\rho_0} 
\equiv
-3\,\gamma\,\frac{\delta r}{r_0} \ .
\label{delp}
\ee 
We stress once more that the quantities carrying the lower index $0$ represent the parameters of the star in the unperturbed state and they obey Eq.~\eqref{acczero}. For $\d m_0$, which is assumed to remain unchanged during the process, the lower index is used to identify the proper mass which is different from the ADM-like mass. 

After applying the perturbation, the equation of motion for the shell~\eqref{acc} becomes 
\be
\d m_0 \,\ddot{r} = -\left(1+\frac{p}{\rho}\right) V'\,\d m_0 - 4\,\pi \, r^2\, \d p \ ,
\label{accdm}
\ee
where, after multiplying the entire equation by the factor $\d m_0$, we have used that $\d m_0 = 4\,\pi \, r^2\,\rho\, \d r$ in the last term. For the matter inside the star following an adiabatic equation of state, this can be rewritten by using Eq.~\eqref{adiabatic} as
\be
\d m_0 \,\ddot{r} = -\left(1+p_0\,\frac{\rho^{\gamma-1}}{\rho_0^{\gamma}}\right) V'\,\d m_0 - 4\,\pi \, r^2\, \d p \ . 
\label{accdm}
\ee

The only remaining unknown is how the potential $V$, more exactly its first derivative, changes as a result of a homologous perturbation. 
Only approximate expressions for the bootstrapped Newtonian potential can be calculated in this framework and the functional dependence on the radius changes not just for different density profiles, but it also varies for different compactness regimes. 
In the following sections two different cases will be investigated separately starting with the simpler stars of uniform density, which will be followed by bootstrapped Newtonian polytropes. 
In both instances the analysis will be limited to stars of small or intermediate compactness. As it shall be clarified in the subsequent section, a good approximation for the intermediate compactness regime upper limit is $X<0.5$. 

\section{Homogeneous stars}
\label{sec:hom}
\setcounter{equation}{0}
We start by investigating stars of homogeneous density
\be
\rho
=
\frac{3\, M_0}{4\,\pi\, R^3}\, 
\Theta(R-r)
\ ,
\label{HomDens}
\ee
where $\Theta$ is the Heaviside step function, and
\be
M_0
=
4\,\pi
\int_0^R
r^2\,\d r\,\rho(r)
\ ,
\ee
represents the proper mass of the star, which can be written in terms of the ADM-like mass~\cite{adm} of the same star as shown in~\cite{Casadio:2019cux,Casadio:2019pli}. The relation between the two will be given shortly. 

The gravitational potential inside the star can be obtained using Eq.~\eqref{EOMVq1} along with some boundary conditions in the centre and across the surface $r=R$ of the star. The potential and its first derivative must be smooth across the star's surface and must match the potential from the outer vacuum, where $\rho=p=0$, region in which Eq.~\eqref{eqP} is trivially satisfied and Eq.~\eqref{EOMVq1} reads
\be
\triangle V
=
\frac{2 \left(V'\right)^2}{1-4\,V}
\ .
\label{EOMV0}
\ee

This equation is exactly solved by
\be
V_{\rm out}
=
\frac{1}{4}
\left[
1-\left(1+\frac{6\,\gn\,M}{r}\right)^{2/3}
\right]
\ ,
\label{sol0}
\ee
where, in an intermediary step, the integration constants were fixed by requiring for the usual Newtonian behaviour in terms of the ADM-like mass $M$ to be recovered for large $r$.  

More specifically, the interior solutions must satisfy the regularity condition in the centre
$
V_{\rm in}'(0)=0
\label{b0}
$,
while across the object's surface we must have 
$
V_{\rm in}(R)
=
V_{\rm out}(R)
\equiv V_{\rm R}
$,
respectively 
$
V'_{\rm in}(R)
=
V'_{\rm out}(R)
\equiv
V'_{\rm R}
$,
where we defined $V_{\rm in}=V(0\le r\le R)$ and $V_{\rm out}=V(R\le r)$.

The absence of a Buchdahl limit means that bootstrapped Newtonian sources can have an arbitrarily large compactness as shown in previous works~\cite{Casadio:2018qeh, Casadio:2019cux}, and it must be clarified what is meant by small or intermediate compactness, respectively what represents the high compactness regime. 
A Newtonian argument can be used to define the horizon as the star radius for which the escape velocity of test particles from its surface equals the speed of light
\be
2\, V_{\rm in}(\rh=R)
=
2\, V_{\rm out}(\rh=R)
=
-1
\ .
\ee
Even without an interior solution for the potential, using the above expression~\eqref{sol0}, one finds the compactness limit where black holes are formed to be $X \simeq 0.69$ and $\rh \simeq R \simeq 1.43\, \gn\,M$.
For larger compactness values the horizon radius will always be located outside the source, somewhere in the
outer potential from Eq.~\eqref{sol0}. 
The regime of compactness around this value or above is considered to be high compactness, while $X \lesssim 0.5$ represents intermediate and (as $X$ decreases further) low compactness. 

While it is impossible to find analytic solutions for the bootstrapped Newtonian potential in the interior of these objects, reliable approximate solutions can be obtained in the low and intermediate compactness regimes by starting from a series expansion of the potential around $r=0$. It can easily be shown that the regularity condition in the centre requires for all odd order terms in $r$ from the Taylor series to vanish~\cite{Casadio:2019cux}. The the conservation equation~\eqref{eqP} along with the remaining two boundary conditions at $r=R$, where the potential and its first derivative must be continuous across the star's surface, are them used to find the approximate solution for the potential 
\be
\label{Vins}
V_{\rm in}
=
\frac{\left[\left(1+6\,X\right)^{1/3}-1\right]
+2\,X\left(r^2/R^2-4\right)}
{4\,\left(1+6\,X\right)^{1/3}}
\ . 
\label{Vintermediate}
\ee
In the process, the constraints at the boundary also result in another useful expression which relates the proper and ADM-like masses
\be
M_0
=
\frac{M\,e^{-\frac{X}{2\left(1+6\,X\right)^{1/3}}}}
{\left(1+6\,X\right)^{1/3}}
\ .
\label{M0}
\ee
The readers interested in dwelling deeper into the subject are referred to the works cited earlier in this section~\cite{Casadio:2019cux,Casadio:2019pli}. 

The first derivative of this approximate potential equals to
\be
 V'_{\rm in}
=
\frac{X\,r}
{R^2\,\left(1+6\,X\right)^{1/3}}
\equiv 
V_c\, r
\ ,
\label{V1intermediate}
\ee
where all the factors not dependent on the radius were included in $V_c$ for the convenience of simplifying the following equations. 

With this, Eq. \eqref{accdm} for the star in the perturbed state becomes
\be
\d m_0 \,\frac{\d^2}{\d t^2}\left[r_0\left(1+ \frac{\delta r}{r_0}\right)\right] 
&=&
-\left[1+\frac{p_0}{\rho_0} \left(1+ \frac{\delta \rho}{\rho_0}\right) ^{\gamma-1}\right] V_c \left[r_0\left(1+ \frac{\delta r}{r_0}\right)\right]  \,\d m_0 
\nonumber
\\
& & -
4\,\pi \, \left[r_0\left(1+ \frac{\delta r}{r_0}\right)\right] ^2\, \d \left[p_0\left(1+ \frac{\delta p}{p_0}\right)\right]  \ .
\label{accdmp_a}
\ee
Equations~\eqref{delrho} and \eqref{delp} can be used to express $\delta \rho/\rho_0$ and $\delta p/p_0$ in terms of $\delta r/r_0$. Furthermore, after Taylor expanding and keeping only the first order terms in $\delta r$ one can use Eq.~\eqref{acczero} for the unperturbed quantities 
\be
-\left(1+\frac{p_0}{\rho_0}\right) V_c\,r_0\,\d m_0 - 4\,\pi \, r_0^2\, \d p_0 =0 \ ,
\label{acczerov1}
\ee
to cancel the zero order terms in $\delta r$. After performing all these algebraic operations, the equation above becomes
\be
\d m_0 \,\ddot{\delta r} 
=
-\d m_0 \left[ 1+\frac{p_0}{\rho_0}\left(4-3\,\gamma\right)\right] V_c\,\delta r
-
4\,\pi \, r_0\left(2-3\,\gamma\right) \d p_0\,\delta r \ .
\label{accdmpsimp}
\ee
One more step can be performed to simplify this equation even more by expressing $\d p_0$ using Eq.~\eqref{acczerov1} and substituting it back into Eq.~\eqref{accdmpsimp} to obtain
\be
\d m_0 \,\ddot{\delta r} 
=
-\d m_0 \left( 3\,\gamma -1+\frac{2\,p_0}{\rho_0}\right) V_c\,\delta r
\ .
\ee
After inserting the expression for $V_c$ and dividing both sides by $\d m_0 $ this becomes 
\be
\ddot{\delta r} 
=
-\frac{X \left[ \left(3\,\gamma -1\right)\rho_0+2\,p_0\right]}
{R^2\,\left(1+6\,X\right)^{1/3}\,\rho_0}
\,\delta r
\ .
\label{HO}
\ee
The solution to this differential equation is of the type 
\be
\delta r = C_+\, e^{i\,\omega\, t} +C_-\, e^{-i\,\omega\, t}
\ ,
\label{HOsol}
\ee
with
\be
\omega=\sqrt{\frac{X \left[ \left(3\,\gamma -1\right)\rho_0+2\,p_0\right]}
{R^2\,\left(1+6\,X\right)^{1/3}\rho_0}}
\ .
\label{omega}
\ee
 
 If the quantity under the square root is positive the solution represents an oscillatory motion and the star is dynamically stable, while if the same quantity is negative one term decays in time, but the other term increases exponentially making the star dynamically unstable. 
The adiabatic index $\gamma$ is equal to $5/3$ for stars made out of non-relativistic gas and it decreases towards $4/3$ as the gas becomes relativistic. In order to gain a broader understanding of the bootstrapped Newtonian stars, we allow the adiabatic index to vary in the range $1\le \gamma \le 2$, both for the homogeneous and polytropic cases. We see from the expression for $\omega$ above, that for the entire range of $\gamma$ taken into consideration the quantity under the square root is always positive. This means that, in response to homologous adiabatic perturbations, homogeneous bootstrapped Newtonian stars have stable oscillatory behaviours. 

\section{Polytropic stars}
\label{sec:polytropes}
\setcounter{equation}{0}
Polytropic stars are characterised by the polytropic equation of state
\be
p(r)
=
\kappa\, \rho^n(r)
\ ,
\label{eosB}
\ee
where $n>1$ represents the polytropic index and $\kappa$ is a constant of proportionality~\cite{horendt}.  
In order to gain a good understanding for a wide variety of stars, the polytropic index is varied in the range $1<n\leq3$~\footnote{An alternative definition for the polytropic index exists, in which the exponent from Eq.~\eqref{eosB} is written in the form $(n+1)/n$, with the correspondingly adjusted ranges for this parameter.}, with the values of $n$ in the upper half of this interval corresponding to neutron stars, while the lower half of the interval covers a variety of other types of stars from white dwarfs to main sequence stars.

In the case of bootstrapped Newtonian polytropic stars, after imposing the available (boundary) constraints, a numerical solution for the density profile that resembles closely to a Gaussian density profile is found~\cite{Casadio:2020kbc}. Under these circumstances one can start by approximating the density profile for the self-gravitating object as
\be
\rho
=
\left\{
\begin{array}{lr}
\strut\displaystyle
\rho_c\,e^{-\frac{r^2}{b^2\,R^2}} 
\ ,
&
r\leq R
\\
\\
\strut\displaystyle
0 \,
\ ,
&
r>R
\ .
\end{array}
\right.
\label{rho_gaussian}
\ee
In order for the density to vanish outside the star we impose a step-like discontinuity at $r=R$. 
This discontinuity is incompatible with the constraint of a vanishing pressure at the surface.
The issue can be mitigated by assuming a (finite) central density and a Gaussian width $b$ such that
 the density on the surface is negligible when compared to the central density and this was shown to represent a good approximation in Ref.~\cite{Casadio:2020kbc}. For this reason, the range for the Gaussian width is limited to $b <1$. A value of $b$ close to one already means that the density near the surface is large from the perspective of the initial assumption necessary for the above approximation. However, considering that the profile is an approximation and a larger value of the parameter $b$ flattens out the density profile, we allow it in order to include a larger parameter space. Another important reason is that it lets us check whether in the limit in which the density profile becomes flat the results from the previous section are reproduced. One way of looking at this slight discontinuity is by considering the surface of the object to be covered by a thin solid crust and a tension in this crust is what balances the non-vanishing pressure. 
A very detailed analysis of bootstrapped Newtonian polytropes, including the approximate expressions for the density profile and gravitational potential that are presented next, can be found in Ref.~\cite{Casadio:2020kbc}.

After some algebra similar to the one detailed in Section~\ref{sec:hom} for homogeneous stars, which involves making use of the regularity condition in the centre and the two boundary conditions at the surface of the star, one can write de density profile of the star as
\be
\rho
=
\frac{3\,n\,X\,e^{(n+1-r^2/R^2)/b^2}\left[\frac{2\,n}{b^2}\left(1+6\,X\right)^{1/3}-X\right]}
{2\,\pi\,\,b^2\,\gn\,R^2\left[e^{n/b^2}\,X+e^{1/b^2}\left(\frac{2\,n}{b^2}\left(1+6\,X\right)^{1/3}-X\right)\right]^2}
\ ,
\label{rhoGauss}
\ee 
while the gravitational potential is given by
\be
V
=
%&\!\!=\!\!&
\frac{n}{(n-1)}
\ln\! \left[\frac{\frac{2\,n}{b^2}(1+6\,X)^{1/3}}{\frac{2\,n}{b^2}(1+6\,X)^{1/3}\!+X\left(e^{(n-1)(1-r^2/R^2)/b^2}\!-\!1\right)}\right]
%\nonumber
%\\
%&&
+
\frac{1}{4}
\left[
1-\left(1+6\,X\right)^{2/3}
\right]
\ .
\label{VRfinal} 
\ee
Starting from the equilibrium density, pressure and gravitational potential, a similar derivation as in the previous section can be followed for bootstrapped Newtonian polytropic stars. The only caveat is that (at least) the intermediary expressions are very complicated and too cumbersome to be displayed herein. However, softwares such as {\it Wolfram Mathematica} \cite{mathematica} can handle them easily. After going through the same steps as in the previous section, the equation for the variation in time of $\delta r$ can be expressed in the form
\be
\ddot{\delta r} 
=
-f(r, X, R,\gamma, n, b)\,\delta r \equiv -f(r)\,\delta r
\ ,
\label{HOa}
\ee
with 
\be
f(r)
&=&
\frac{2\,n\,X\,e^{\frac{(n-1)(R^2-r^2)}{b^2\,R^2}}}
{b^2\,R^2\left(2\,n\,Y-b^2\,X\right)
\left[ b^2\,X\left(e^{\frac{r^2+n\,R^2}{b^2\,R^2}}-e^{\frac{n\,r^2+R^2}{b^2\,R^2}}\right)
+2\,n\, e^{\frac{n\,r^2+R^2}{b^2\,R^2}}Y
\right]}
\nonumber
\\
&&
\left\{
2\,b^2\,(3\,\gamma-1)\,n\,Y\,e^{\frac{n\,r^2+R^2}{b^2\,R^2}}
-b^4\,X\left[(3\,\gamma-1)\,e^{\frac{n\,r^2+R^2}{b^2\,R^2}}-2\,e^{\frac{r^2+n\,R^2}{b^2\,R^2}}\right]
\right.
\nonumber
\\
&&
\left.
-
2\left(n-1\right)\left(2\,n\,Y-b^2\,X\right)e^{\frac{n\,r^2+R^2}{b^2\,R^2}}\,
\frac{r^2}{R^2}
\right\}
\ ,
\label{fX}
\ee
where, in order to shorten the expression, we have used $Y\equiv (1+6\,X)^{1/3}$. 

\begin{figure}[tp]
\centering
\includegraphics[width=5.3cm]{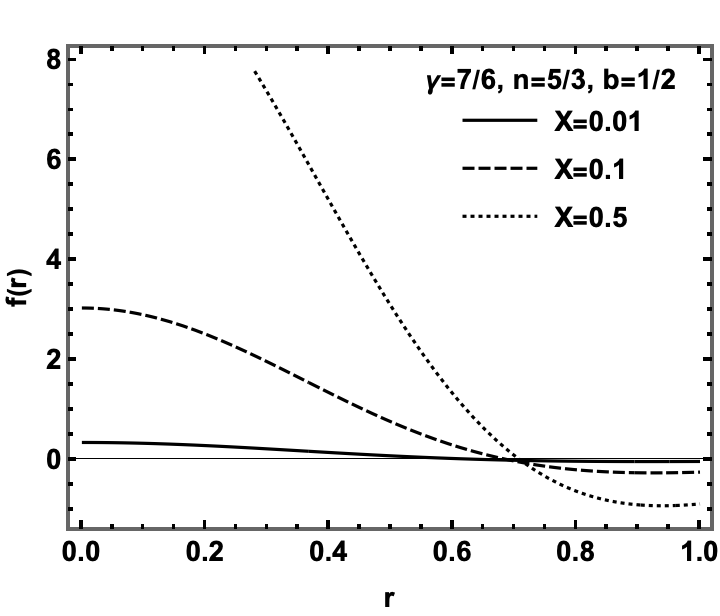}
\includegraphics[width=5.3cm]{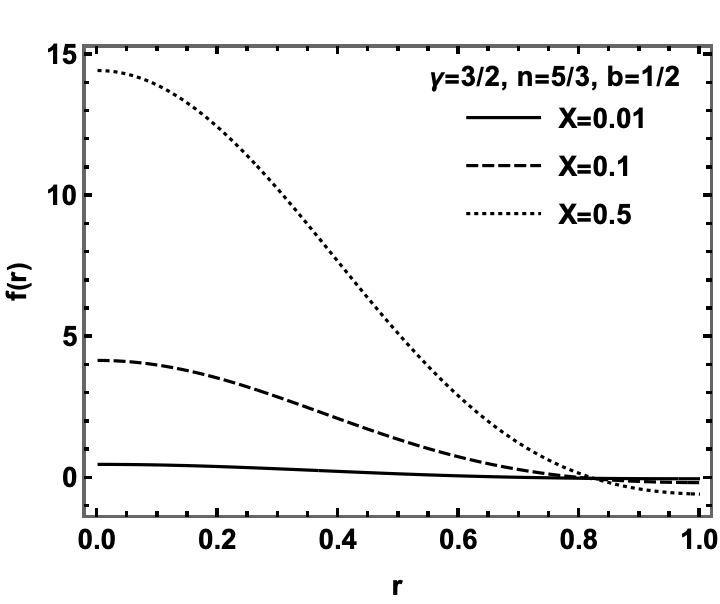}
\includegraphics[width=5.3cm]{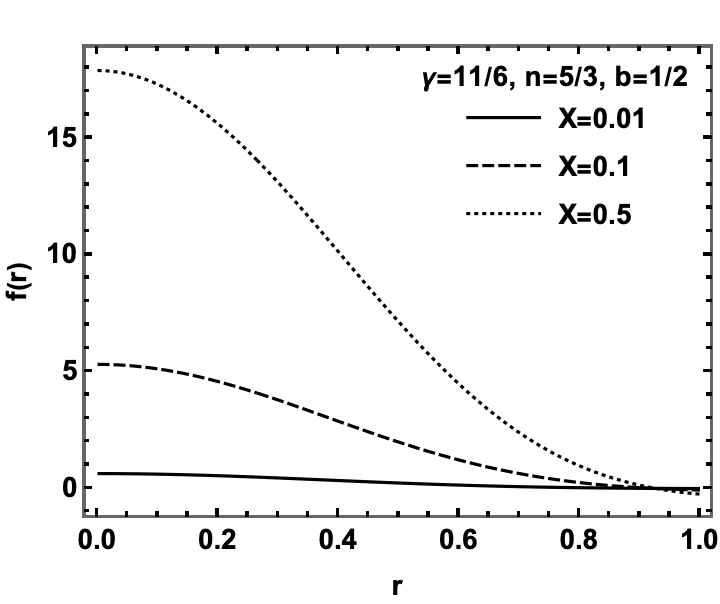}
\\
\includegraphics[width=5.3cm]{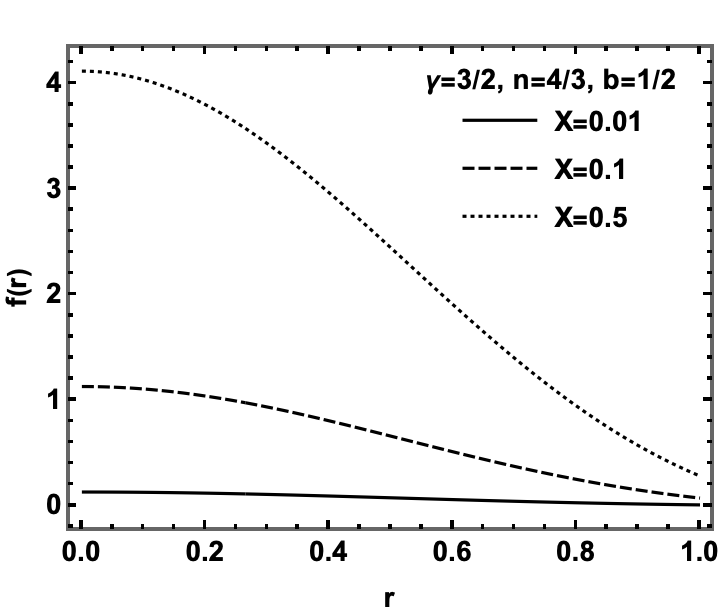}
\includegraphics[width=5.3cm]{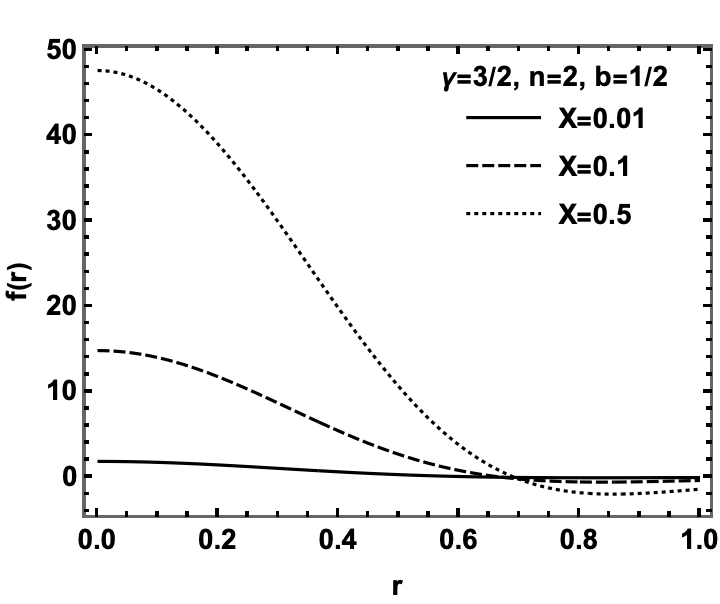}
\includegraphics[width=5.3cm]{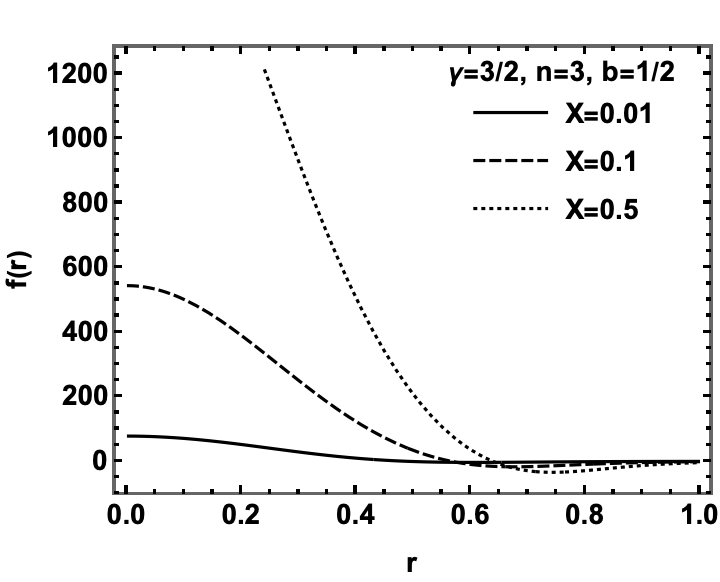}
\includegraphics[width=5.3cm]{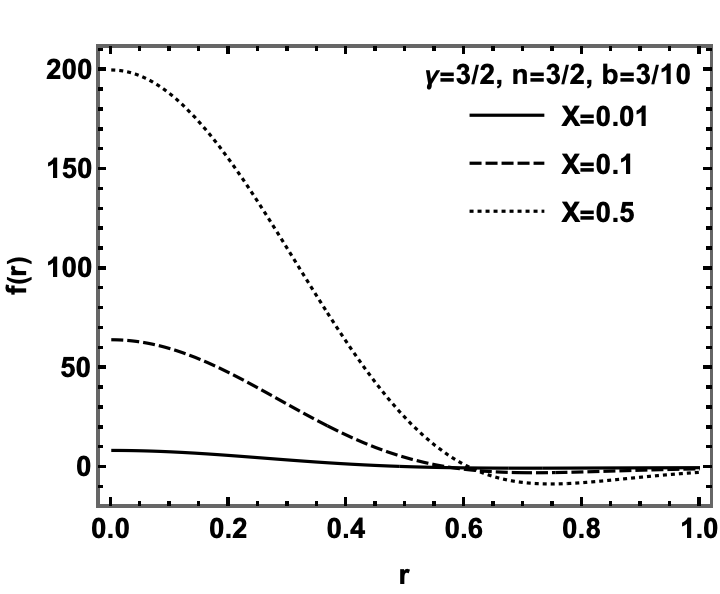}
\includegraphics[width=5.3cm]{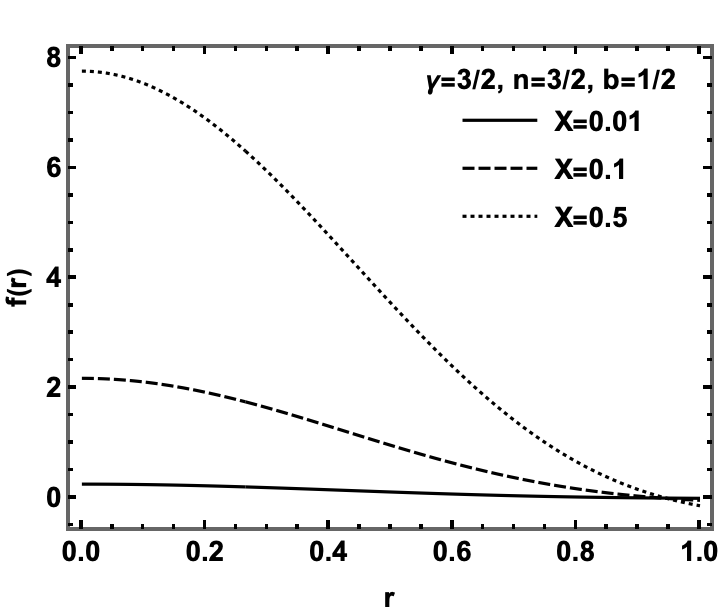}
\includegraphics[width=5.3cm]{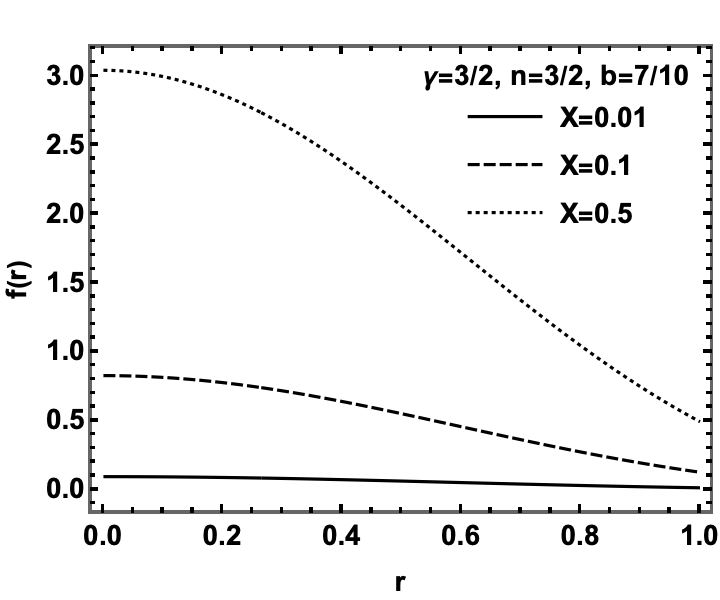}
\caption{Plots of $f(r)$ as a function of $r$ for $R=1$. \underline{Top panels}: the adiabatic index increases from left to right. \underline{Middle panels}: the polytropic index $n$ increases from left to right. \underline{Bottom panels}: the Gaussian width $b$ varies, increasing from left to right. Note the different ranges on the vertical axis.}
\label{Fig1}
\end{figure}

Once more, the solution can be written in the form from Eq.~\eqref{HOsol}, with $\omega=\sqrt{f(r)}$. Depending on the sign of the function $f(r)\equiv f(r, X, R,\gamma, n, b)$ we can have a dynamically stable solution for $f(r)>0$, respectively an unstable one when $f(r)<0$, as explained in the previous section. We are mostly interested in finding the parameter space for solutions that have a stable behaviour. The fraction on the top row of Eq.~\eqref{fX} is always positive for our ranges of parameters. In particular, the relevant constraints on the parameters that lead to this conclusion are $b<1$, $n>1$, respectively $X<0.5$. However, the remaining term can change sign and we cannot obtain clear analytical ranges for the free parameters in which the expression is either positive or negative. This term is a function of $r/R$ only, with no additional dependence on the star radius. 
For this reason, without loosing any generality, in Fig.~\ref{Fig1} we plot the function $f(r)$ as a function of $r$, with $R=1$, for three values of the compactness from low to intermediate and various sets of the remaining parameters. 
Taking $R=1$ does not result in any loss of generality because the only dimension-full quantity in Eq.~\eqref{fX} is an overall factor of $X/R^2\simeq \gn M/R^3$ which varies in magnitude depending on the compactness of the star and its mass but does not chance sign, therefore it does not impact the current analysis.
We first remark that, in many of the plots shown in Fig.~\ref{Fig1}, $f(r)$ changes sign for values of $r$ somewhere between $0$ and $R$. In order for a solution to be oscillatory, this function must be positive throughout the entire interval $0\leq r\leq R$. 
As $r$ increases towards the surface of the star, in most cases, $f(r)$ decreases continuously and in many instances it also becomes negative. There are a few exceptions in which after decreasing to a minimum negative value, the function starts to slowly increase again near the surface, however, for our choice of parameters it remains below zero.  
Having in mind the majority of the cases for which the function continues to decrease all the way to the star's surface, it is interesting to analyse the parameter space for which the function is positive at $r=R$. 
This is a fairly good indicator for finding the ranges of parameters for which the stars are dynamical stable. 
Considering that in some cases the function has a minimum and then starts to increase again, for any set of parameters, the stability of the star needs to be evaluated in its entire volume from the center to the surface. 

At $r=R$, the expression from~\eqref{fX} simplifies to
\be
f(R)
=
\frac{X\left\{
2\,b^2
\left[
(n-1)\,X+(3\,\gamma-1)\,n\,Y
\right]
-
4\,n\,(n-1)\,Y
-
3\,b^4\,X (\gamma-1)
\right\}
}
{b^2\,R^2\,Y\left(2\,n\,Y-b^2\,X\right)
}
\ .
\label{fXR}
\ee
\begin{figure}[tp]
\centering
\includegraphics[width=5.3cm]{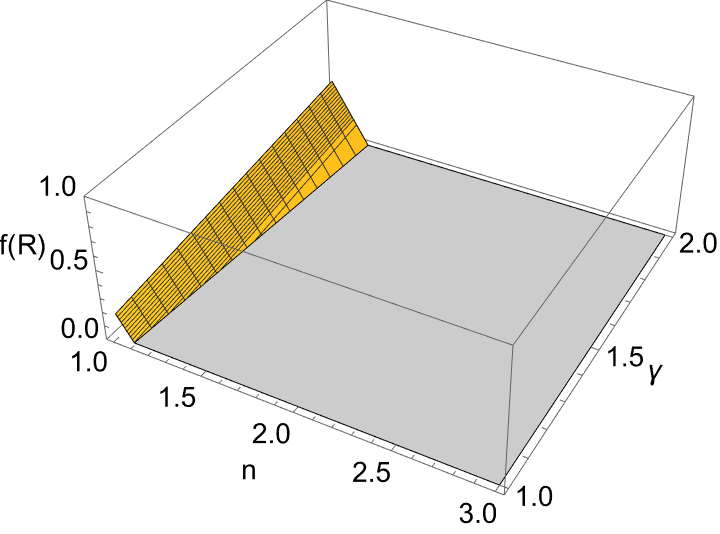}
\includegraphics[width=5.3cm]{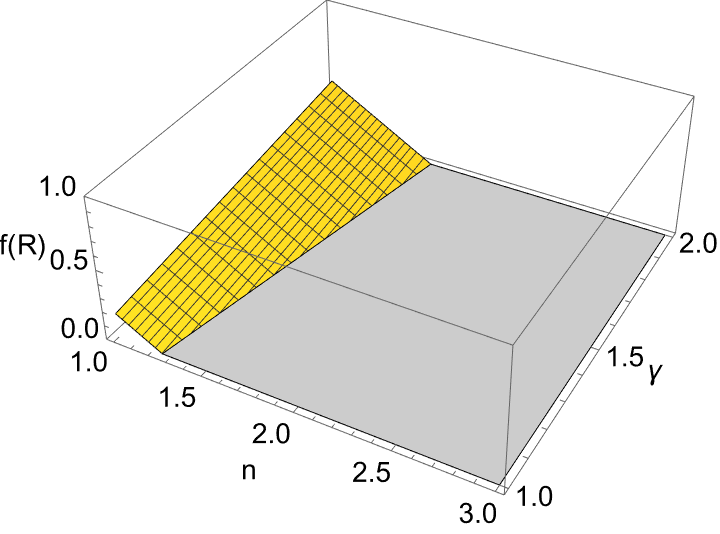}
\includegraphics[width=5.3cm]{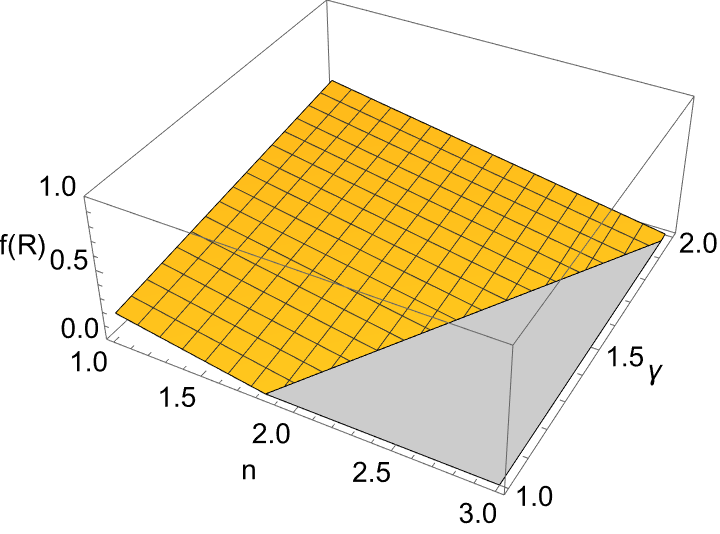}
\\
\includegraphics[width=5.3cm]{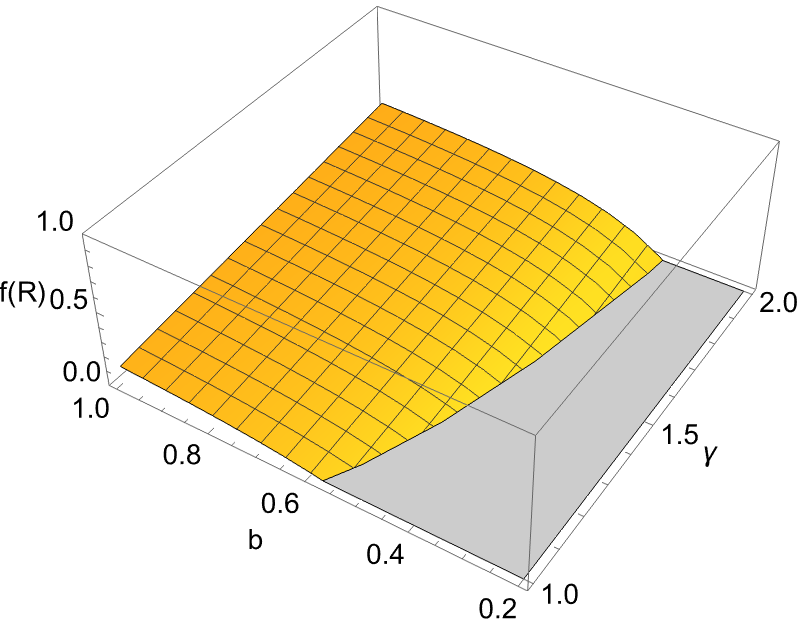}
\includegraphics[width=5.3cm]{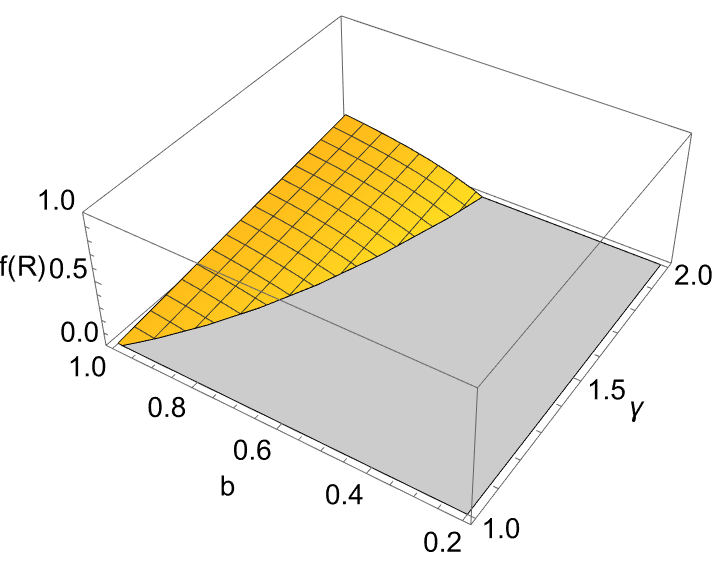}
\includegraphics[width=5.3cm]{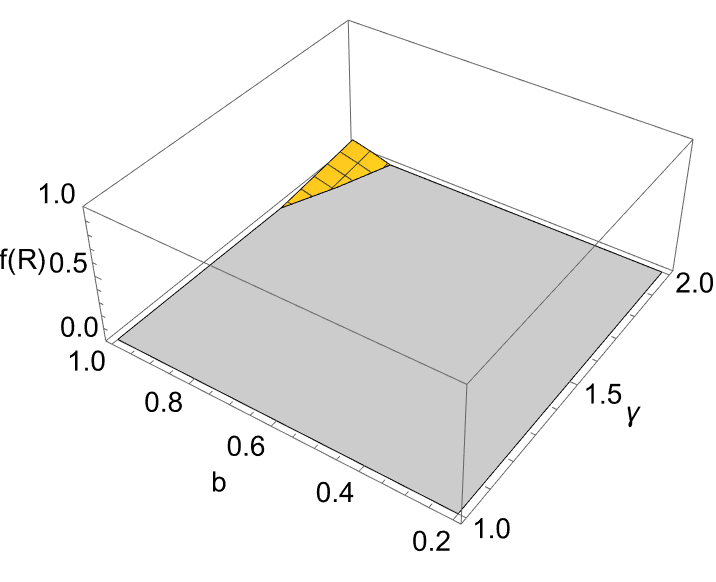}
\\
\includegraphics[width=5.3cm]{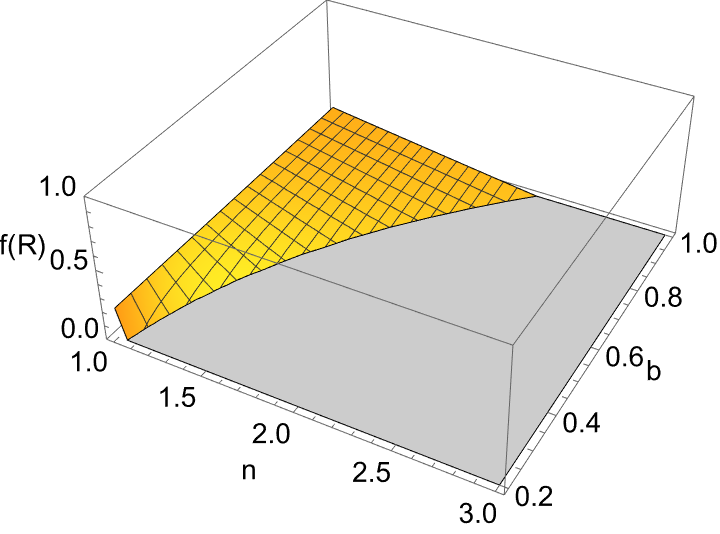}
\includegraphics[width=5.3cm]{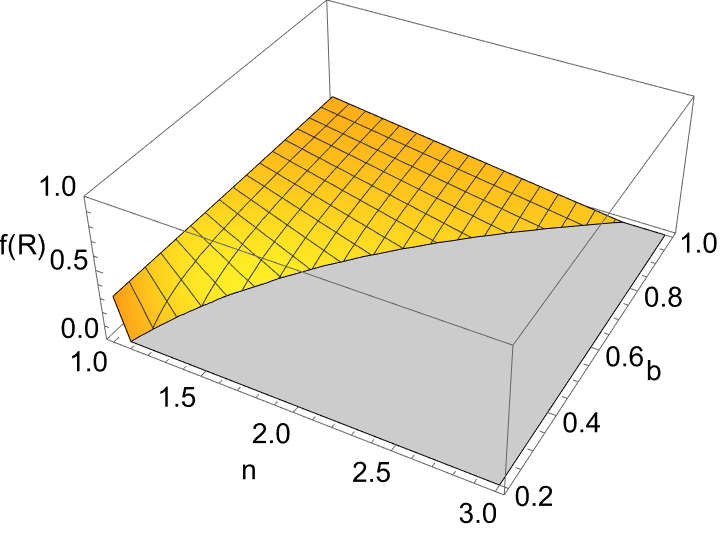}
\includegraphics[width=5.3cm]{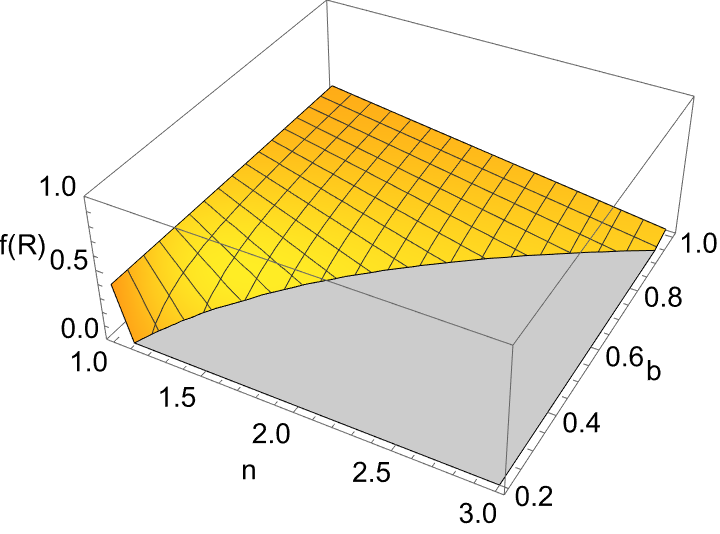}
\caption{\underline{Top panels}: 3D plots of $f(R)$ for $b=3/10$ (left), $b=1/2$ (center) and $b=9/10$ (right). \underline{Middle panels}: 3D plots of f(R) for $n=4/3$ (left), $n=2$ (center) and $n=3$ (right). \underline{Bottom panels}: 3D plots of f(R) for $\gamma=7/6$ (left), $\gamma=3/2$ (center) and $\gamma=11/6$.}
\label{Fig2}
\end{figure}
In the ranges of interest for the free parameters the denominator of $f(R)$ is always positive and the overall sign is given by the numerator, more exactly the expression between the curly brackets. 
Fig.~\ref{Fig2} shows a set of three dimensional plots for stars with a compactness value of $X=0.1$. We are only interested in the parameter ranges for which Eq.~\eqref{fXR} is positive. It was argued earlier that we can use $R=1$ without loosing any generality. In each row of plots one of the three remaining free parameters $\gamma$, $n$, respectively $b$ takes (increasing) discrete values, while the other two span their entire allowed ranges. The plots show the regions in which $f(R)$ is positive. 
The top panels show that as the Gaussian width $b$ increases, the ranges of $\gamma$ and $n$ for which $f(R)$ is positive also increase. 
This represents the expected behaviour, because in the large $b$ limit the density profile flattens out and ideally should reproduce the homogeneous density profile discussed in detail in Section~\ref{sec:hom}. 
For uniform densities it was shown that the the quantity under the square root of Eq.~\eqref{omega}, which is equivalent to $f(r)$, is always positive making these stars always stable. On the opposite regime, for small $b$ values, the parameter space is very constrained and only polytropes with a very small index $n$ can be stable. 
The middle panels show that as the value of the polytropic index $n$ increases, larger values of the Gaussian width $b$ are required in order for the solution to correspond to a star with stable oscillatory behaviour. Even if the dependence on the adiabatic index value $\gamma$ is smaller, for a constant index $n$, as $b$ decreases so does the interval of $\gamma$ values which leads to stable solutions. 
For $n\gtrsim 2$ only the larger values from the allowed range of the parameter $b$ correspond to stable oscillatory solutions.
A similar analysis is performed in the plots from the bottom row, in which three values covering the allowed range are chosen for the adiabatic index $\gamma$. As the value of $\gamma$ increases, wider ranges of values of the Gaussian width $b$ and polytropic index $n$ lead to stable solutions. 

\begin{figure}[tp]
\centering
\includegraphics[width=5.2cm]{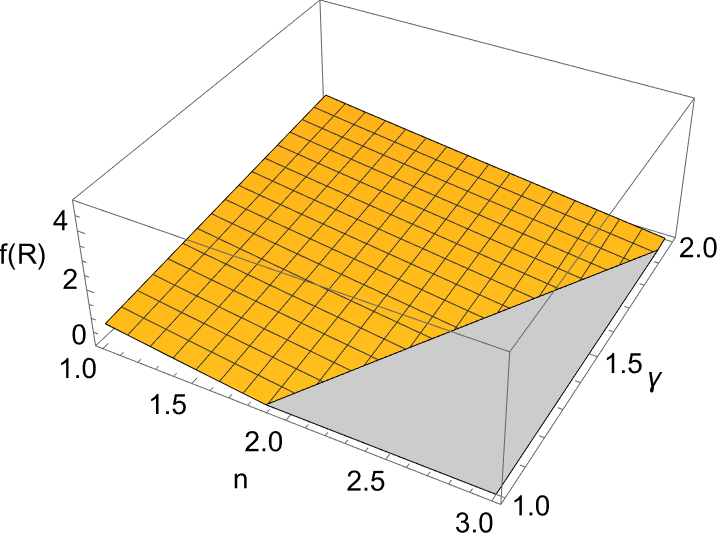}
\includegraphics[width=5.2cm]{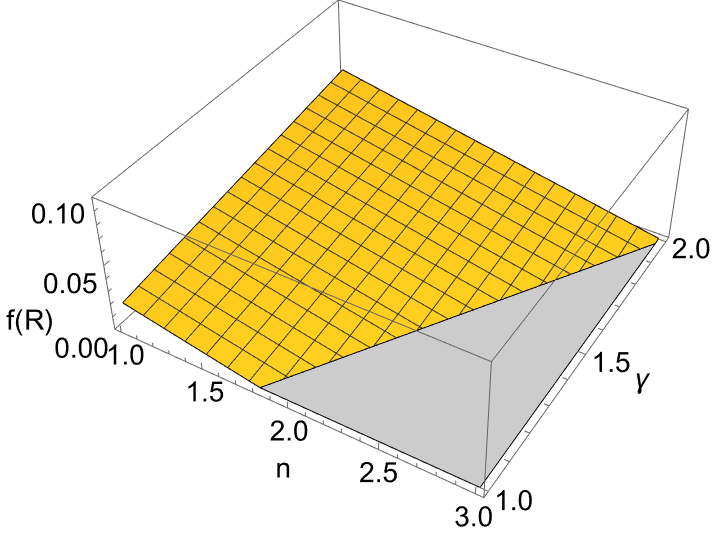}
\includegraphics[width=5.5cm]{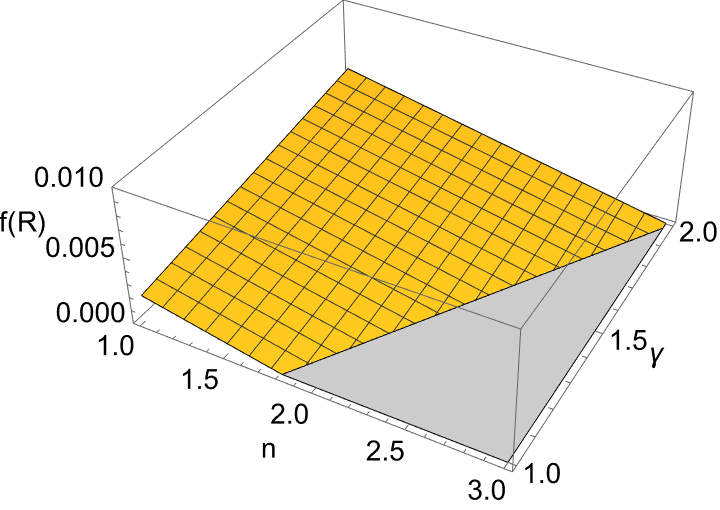}
\caption{3D plots of $f(R)$ for  $b=9/10$ and $X=0.5$ (left), $X=10^{-2}$ (middle) and $X=10^{-3}$ (right).}
\label{Fig3}
\end{figure}

An extensive numerical analysis, similar to the one presented in Fig.~\ref{Fig2}, was performed for $10^{-3}~\leq~X~\leq~0.5$ and it showed that the value of the compactness has a negligible impact on the stability of bootstrapped Newtonian stars, when all other parameters are kept constant. As an example, Fig.~\ref{Fig3} shows a set of three dimensional plots corresponding to $b=9/10$ for three more values of the compactness. Along with the corresponding plot from Fig.~\ref{Fig2}, these show that $f(R)$ is positive for approximately the same combinations of adiabatic and polytropic indices, regardless of the compactness. Similar results are obtained when varying $X$ for the other cases presented in Fig.~\ref{Fig2}.
Of course, that does not imply that stars of very different values of the compactness will be characterised by the same indices. 
It only allows us to simplify the analysis by picking a certain compactness value and focusing on the remaining parameters, which are responsible for determining the dynamical stability of the bootstrapped Newtonian polytropic stars. 

While it was found that stars of constant density are generally dynamically stable, bootstrapped Newtonian polytropic stars are unstable for a large section of the parameter space. In the case of dynamically unstable solutions, small perturbations such as those considered here will continue to increase in size to a point where this analysis fails, and this happens over a characteristic time period on the order of $t_{\rm inst}\simeq 1/\sqrt{-f(r)}$. The negative sign under the square root was inserted because unstable solutions correspond to $f(r)<0$.

To summarise, depending on the values of all of the parameters from Eq.~\eqref{fXR}, they can lead to dynamically stable or unstable solutions. However, the strongest dependence is contained in the shape of the density profile. 

\section{Conclusions}
\label{sec:conc}
\setcounter{equation}{0}

In this article we analysed the dynamical stability of bootstrapped Newtonian stars following some homologous density perturbations which are also assumed to be adiabatic. 
The analysis was limited to stars of small or intermediate compactness, even if in the absence of a Buchdahl limit pressure can also sustain objects of large compactness in equilibrium. However, due to difficulties in finding approximate solutions for the potential inside high compactness sources, those cases are deferred for a future work. 

Two separate cases were taken into consideration. In Section~\ref{sec:hom}, stars with uniform densities were shown to be stable to the appearance of some adiabatic density perturbations such as the ones discussed above. These perturbations result in an oscillatory behaviour of the stars, regardless of their (low or intermediate) compactness $X$, density value $\rho_0$, or adiabatic index $\gamma$. 

In Section~\ref{sec:polytropes} we analysed polytropes, more exactly polytropic stars with density profiles approximated by Gaussian-like functions as in Eq.~\eqref{rho_gaussian}. In this case it was shown that, depending on the parameters characterising these objects,  small perturbations can lead to both stable oscillating solutions and unstable solutions. For the unstable cases, the instabilities appeared primarily in the outer layers of the star. Therefore, the next step was to evaluate the behaviour of the outermost layers of the bootstrapped Newtonian polytropic stars and conclude that when these layers had a stable oscillatory behaviour, it is highly likely for the entire object to be stable. It must also be emphasised  that in order to be sure that a star characterised by a certain set of parameters is stable, one should analyse its behaviour throughout the entire volume. 

When evaluating the stability of the outer layers, described by the function from Eq.~\eqref{fXR}, an extensive numerical analysis showed that the value of the compactness $X$ has a fairly negligible effect on the stability of the star when all other parameters are kept constant.  
The next important conclusion is that wider (flatter) Gaussian density profiles lead to stable solutions for greater ranges of the other remaining parameters such as the adiabatic or polytroopic indices. This finding is consistent with the results from Section~\ref{sec:hom}, since in the large $b$ limit the density profile resembles a uniform density profile.
When evaluating the dependence on the polytropic index, the parameter space for stable solutions is larger for smaller values of $n$. According to Fig.~\ref{Fig2} bootstrapped Newtonian stars with $n\gtrsim 2$ (corresponding to neutron stars) can only be stable for larger values of the Gaussian width, i.e. for flatter density profiles.
Finally, the opposite is true when considering the dependence on the adiabatic index $\gamma$: the ranges of $n$ and $b$ that lead to stable solutions increase with $\gamma$. 
For unstable solutions, perturbations increase in size in some characteristic time periods on the order of $t_{\rm inst}\simeq 1/\sqrt{-f(r)}$. 

To summarise, the bootstrapped Newtonian model can lead to stars that are dynamically stable following the appearance of some adiabatic density perturbations. What remains to be investigated is whether the same conclusions are reached for other types of perturbations such as thermal stability (which happens for instance when the star exceeds thermal equilibrium). It will be interesting to find out if the range of parameters for which we have thermal stability for bootstrapped Newtonian stars overlaps with the parameter space in which we found that these stars are dynamically stable. This analysis is deferred to some future works. 

\section*{Acknowledgments}
This research was supported by the Romanian Ministry of Research, Innovation and Digitalization under the Romanian National Core Program LAPLAS VII - contract no. 30N/2023. 
%\clearpage
%
% 
%%%%%%%%%%%%%%%%%%%%%%%%%%%%%%%%%%%%%%%%%%%%%%%%%%%%%%%%%%%%%%%%%
%%%
%%%                     BIBLIOGRAPHY
%%%
%%%%%%%%%%%%%%%%%%%%%%%%%%%%%%%%%%%%%%%%%%%%%%%%%%%%%%%%%%%%%%%%%
%
%

%
\end{document}